\documentclass[prl,showpacs,twocolumn]{revtex4}
\usepackage{amssymb}
\usepackage{amsmath}
\usepackage{amssymb}
\usepackage{graphicx}
\usepackage{amsmath}
\usepackage{colordvi}
\usepackage{bbm}
\usepackage{epsfig}
\usepackage{amsfonts}
\usepackage{color}
\usepackage{float}
\usepackage{latexsym}
\usepackage{mathrsfs}
\usepackage{feynmf}
\def\QED{\leavevmode\unskip\penalty9999 \hbox{}\nobreak\hfill
     \quad\hbox{\leavevmode  \hbox to.77778em{%
               \hfil\vrule   \vbox to.675em%
               {\hrule width.6em\vfil\hrule}\vrule\hfil}}
     \par\vskip3pt}
\def\qed{\leavevmode\unskip\penalty9999 \hbox{}\nobreak\hfill
     \quad\hbox{\leavevmode  \hbox to.77778em{%
               \hfil\vrule   \vbox to.675em%
               {\hrule width.6em\vfil\hrule}\vrule\hfil}}
\par\vskip3pt}
\def\ibb #1{\leavevmode\hbox{\kern.3em\vrule
     height 1.5ex depth -.1ex width .4pt\kern-.3em\rm#1}}

\newcommand{\be}{\begin{equation}}
\newcommand{\ee}{\end{equation}}
\newcommand{\ba}{\begin{array}}
\newcommand{\ea}{\end{array}}
\newcommand{\bqa}{\begin{eqnarray}}
\newcommand{\eqa}{\end{eqnarray}}

\newcommand{\tr}{\mbox{Tr}}

\newcommand{\eins}{\ensuremath{\mathbbm 1}}

\begin{document}

\title{Detecting the concurrence of an unknown state with a single observable}
\author {Zhi-Hao Ma$^{1}$, Zhi-Hua Chen$^{2}$, Jing-Ling Chen$^{3}$}

\affiliation { Department of Mathematics, Shanghai
Jiaotong University, Shanghai, 200240, P. R. China }

\affiliation {Department of Science, Zhijiang college, Zhejiang
University of technology, Hangzhou, 310024, P.R.China}

\affiliation {Theoretical Physics
Division, Chern Institute of Mathematics, Nankai University,
Tianjin, 300071, P.R.China}
\begin{abstract}
While the detection of entanglement has been proved already to be quite
a difficult task, experimental quantification of entanglement is even more challenging.
In this work, we derive an analytical lower bound for the concurrence of a bipartite mix
quantum state in arbitrary dimension. The lower bound is experimentally
implementable in a feasible way, which enabling quantification of entanglement in a broad variety of cases.
\end{abstract}


\maketitle

Entanglement is a distinctive feature of quantum mechanics \cite{Horodecki09,Guhne09}, and an indispensable ingredient in various kinds of quantum information processing applications vary from quantum cryptography \cite{Ekert91} and quantum teleportation \cite{Bennett93} to
measurement-based quantum computing \cite{BRaussendorf01}.

The use of entanglement as a resource not only bears the question
of how it can be detected, but also how it can be quantified. For this purpose, several entanglement measures
have been introduced, one of the most prominent of which
is the concurrence \cite{Wootters98}.

However, calculation of the concurrence is a formidable task as the Hilbert
space dimension is increasing. Good algorithms and
progresses have been obtained concerning lower bounds
\cite{chenp02-Gerjuoy03,Lozinski03,Audenaert01,Mintert04}. Considerable progress is made
in order  to give a purely algebraic lower bound and  experimental verifying  it \cite{Mintert05PR,Fei05,Breuer06,Vicente0708,MintertNature06,Mintert07,Mintert08,Ma09,Augusiak,Ma10,Sargolzahi11}.

We start with a generalized definition of
concurrence for a pure state ${\left\vert \psi \right\rangle }$ in the
tensor product $\mathcal{H}_A \otimes \mathcal{H}_B$ of two Hilbert spaces $\mathcal{H}_A,\mathcal{H}_B$ for systems $A,B$.The concurrence is defined by \begin{equation}C({\left\vert \psi \right\rangle }):=\sqrt{%
2(1-\mbox{Tr}\rho _{A}^{2})}\end{equation} where the reduced density matrix $\rho _{A}$
is obtained by tracing over the subsystem $B$.The concurrence is then extended to mixed states $\rho$ by the convex roof,\begin{equation}
C(\rho )\equiv \min_{\{p_{i},|\psi _{i}\rangle \}}\sum_{i}p_{i}C({\left\vert
\psi _{i}\right\rangle }),
\end{equation}
for all possible ensemble realizations $\rho =\sum_{i}p_{i}|\psi _{i}\rangle
\langle \psi _{i}|$, where $p_{i}\geq 0$ and $\sum_{i}p_{i}=1$. From definition, a
state $\rho $ is \emph{separable} if and only if $C(\rho )=0$.

In this letter, we will show that strong lower bounds of concurrence  can
be derived by exploiting close  relations between
concurrence and a recently introduced detection criteria for bipartite entanglement.
The lower bound we found is not only analytical, but also is physical accessible, i.e.,we can easily use the existing  experiments technology to obtain this bound(e.g., \cite{MintertNature06,Mintert07,Mintert08,Huang09}).

To disclose the connection between concurrence and entanglement detection criteria, let us first review the known criterions. Until now, many possible ways to detect entanglement  have
been proposed. These range from Bell inequalities, entanglement witnesses and spin squeezing inequalities to
entropic inequalities, the measurement of nonlinear properties of the quantum state and the approximation
of positive maps, see\cite{Horodecki09,Guhne09}.

Among so many methods, the  PPT criterion and the realignment criterion are the two distinguished ones, they are most powerful and widely used by quantum information community. The Peres-Horodecki criterion of positivity under partial transpose (PPT criterion)\cite{Peres,HorodeckiPLA96} say that $\rho ^{T_{A}}\geq 0 $ should be satisfied for a
separable state, where $\rho ^{T_{A}}$ stands for a partial transpose with
respect to the subsystem $A$. This criterion is even to be sufficient for $
2\times 2$ and $2\times 3$ bipartite systems. However, for high dimension system, there exists  entanglement states that are positive under partial transpose, the so called bound entanglement states (BES)\cite{Horodeckibound}.

Another complementary operational criterion for separability called the
\emph{realignment} criterion is very strong in detecting many of BES \cite{realignment}. This criterion
states that a realigned version $\mathcal{R}(\rho )$ of $\rho $ should
satisfy $||\mathcal{R}(\rho )||\leq 1$ for any separable state $\rho $. Note that realignment criterion is in some sense `` dual " to the PPT criterion, in the following way: A density operator can be written as
\begin{eqnarray}
\rho &=&\sum_{ijkl}\langle ij|\rho|kl\rangle |ij\rangle \langle kl|\nonumber \\
&=&\sum_{ijkl}\rho_{ij,kl}|i\rangle \langle k|\otimes |j\rangle \langle l|
\end{eqnarray}
A partial transpose with respect to the first system $A$ is
\begin{equation}
\left( \rho^{T_{A}}\right) _{ij,kl}:=\rho_{kj,il}.
\end{equation}%

While the realignment of $\rho$ is defined as:
\begin{equation}
\left( \rho^{R}\right) _{ij,kl}:=\rho_{ik,jl}.
\end{equation}%

Recently, a criterion was introduced, which was strictly stronger than the realignment criterion. It says that, if a state is separable, then the following must hold\cite{Zhang08R}:
\begin{equation}
\|\mathcal{R}(\rho-\rho_{A}\otimes\rho_{B})\|
\leq \sqrt{(1-\mathrm{Tr}\rho_{A}^{2})(1-\mathrm{Tr}\rho_{B}^{2})}]
\end{equation}%

We now derive the main result of this Letter.

{\bf Theorem:} For any $m\otimes n$ $(m\leq n)$ mixed
quantum state $\rho $, the concurrence $C(\rho )$ satisfies the following:
\begin{equation}
C(\rho)\geq \sqrt{\frac{2n}{(n-1)(n+1)^{2}}}\times f(\rho).
\end{equation}

Where $f(\rho):=[\|\mathcal{R}(\rho-\rho_{A}\otimes\rho_{B})\|
-\sqrt{(1-\mathrm{Tr}\rho_{A}^{2})(1-\mathrm{Tr}\rho_{B}^{2})}]$.

\emph{Proof.---} Without loss of generality, we
suppose that a pure $m\otimes n$ $(m\leq n)$ quantum state $\rho:=|\psi\rangle\langle \psi|$ has the standard
Schmidt form
\begin{equation}
{|\psi\rangle }=\sum_{i}\sqrt{\mu_{i}}{|
a_{i}b_{i}\rangle },
\end{equation}%
where $\sqrt{\mu _{i}}$ $(i=1,\ldots m)$ are the Schmidt coefficients, ${%
|a_{i}\rangle }$ and ${| b_{i}\rangle }$ are
orthonormal basis in $\mathcal{H}_{A}$ and $\mathcal{H}_{B}$, respectively.
The two reduced density matrices $\rho_{A}$ and $\rho_{B}$
have the same eigenvalues of $\mu_{i}$. It follows%
\begin{equation}
C^{2}({\rho })=2\Big(1-\sum_{i}\mu_{i}^{2}\Big)%
=4\sum_{i<j}\mu_{i}\mu_{j},  \label{squareconcurrence}
\end{equation}%
which varies smoothly from $0$, for pure product states, to $2(m-1)/m$ for
maximally entangled pure states.

Then we get
\[\begin{array}{lllll}
\|R(\rho-\rho_{A}\otimes\rho_{B})\|\\
=\|[\sum\limits_{i}(\mu_{i}-\mu^{2}_{i})|
a_{i}a_{i}\rangle \langle b_{i}b_{i}|+\sum\limits_{i\neq j}\sqrt{\mu_{i}\mu_{j}}|
a_{i}a_{j}\rangle \langle b_{i}b_{j}|\\
\hspace{0.5cm}-\sum\limits_{i\neq j}\mu_{i}\mu_{j}|a_{i}a_{i}\rangle \langle b_{j}b_{j}|]\|\\
\leq \|\sum\limits_{i}(\mu_{i}-\mu^{2}_{i})\|+\sum\limits_{i\neq j}\sqrt{\mu_{i}\mu_{j}}+\sum\limits_{i\neq j}\mu_{i}\mu_{j}\\
=\sum\limits_{i}(\mu_{i}-\mu^{2}_{i})+\sum\limits_{i\neq j}\sqrt{\mu_{i}\mu_{j}}+\sum\limits_{i\neq j}\mu_{i}\mu_{j}
\end{array}\]

So from Cauchy-Schwarz inequality, we get
\begin{equation}\begin{array}{lllll}
\|R(\rho-\rho_{A}\otimes\rho_{B})\|-[\sqrt{(1-\mathrm{Tr}\rho_{A}^{2})(1-\mathrm{Tr}\rho_{B}^{2})}]\\
\leq 2\sqrt{\mu_{1}\mu_{2}}(1+\sqrt{\mu_{1}\mu_{2}})+2\sqrt{\mu_{1}\mu_{3}}(1+\sqrt{\mu_{1}\mu_{3}})\\
\hspace{0.5cm}+\cdots+2\sqrt{\mu_{n-1}\mu_{n}}(1+\sqrt{\mu_{n-1}\mu_{n}})\\
\leq 2\sqrt{\sum\limits_{i< j}\mu_{i}\mu_{j}}\sqrt{\sum\limits_{i< j}(1+\sqrt{\mu_{i}\mu_{j}})^{2}}\\
\leq 2\sqrt{\sum\limits_{i< j}\mu_{i}\mu_{j}}\sqrt{\sum\limits_{i< j}(1+2\sqrt{\mu_{i}\mu_{j}})+\sum\limits_{i< j}(\mu_{i}\mu_{j})}\\
\leq 2\sqrt{\sum\limits_{i< j}\mu_{i}\mu_{j}}\sqrt{\sum\limits_{i< j}(1+(\mu_{i}+\mu_{j}))+\frac{n-1}{2n}}\\
=2\sqrt{\sum\limits_{i< j}\mu_{i}\mu_{j}}\sqrt{\frac{n(n-1)}{2}+(n-1)+\frac{n-1}{2n}}\\
=\sqrt{\frac{(n-1)(n+1)^{2}}{2n}}\times C(\rho)
\end{array}\nonumber
\end{equation}

Now, we will show that the inequality (7) also holds for mix state.Assume we have found the optimal decomposition $\sum_{i}p_{i}\rho_{i}$ for $\rho $ to achieve the infimum of $C(\rho )$, where $\rho_{i}$ are pure
state density matrices. Then $C(\rho )=\sum_{i}p_{i}C(\rho_{i})$ by
definition. Now, we need to prove that after mixture, the bound  will not become bigger, i.e.,$f(\sum_{i}p_{i}\rho_{i})\leq \sum_{i}p_{i}f(\rho_{i})$.From the convex property
of the trace norm, we know that $\|\mathcal{R}(\rho-\rho_{A}\otimes\rho_{B})\|$ is decreasing after convex combination, so the only thing left to prove is that $\sqrt{(1-\mathrm{Tr}\rho_{A}^{2})(1-\mathrm{Tr}\rho_{B}^{2})}$ is increasing after convex combination. It is sufficient to consider the case of $\rho=\frac{1}{2}\rho_{1}+\frac{1}{2}\rho_{2}$. Now, we will prove that
\begin{equation}\begin{array}{lllll}
\sqrt{[1-\frac{1}{4}\tr(\rho_{1A}+\rho_{2A})^{2}][1-\frac{1}{4}\tr(\rho_{1B}+\rho_{2B})^{2}]}\\
\geq \frac{1}{2}[1-\tr(\rho^{2}_{1A})]+\frac{1}{2}[1-\tr(\rho^{2}_{2A})]
\end{array}\end{equation}

For clear, denote $\tr(\rho^{2}_{1A})=\tr(\rho^{2}_{1B}):=x_{1}$, $\tr(\rho^{2}_{2A})=\tr(\rho^{2}_{2B}):=x_{2}$,
$\tr(\rho_{1A}\rho_{2A}):=x_{3}$,$\tr(\rho_{1B}\rho_{2B}):=x_{4}$, then the inequality (10) reduced to prove that the function $F:=-\frac{3}{16}(x_{1}+x_{2})^{2}+\frac{1}{8}(x_{1}+x_{2})(x_{3}+x_{4})+\frac{1}{4}x_{3}x_{4}
-\frac{1}{2}(x_{3}+x_{4})+\frac{1}{2}(x_{1}+x_{2})\geq 0$.To get the minimal value of $F$, we will use the Lagrange multipliers method, and find that $F$ attained its minimal value at the point $(x_{1},x_{2},x_{3},x_{4})=(1,1,1,1)$. And in this case, the minimal value is $F=0$. So we get $F\geq 0$. Theorem is proved.\hfill\rule{1ex}{1ex}

The most prominent feature of this theorem is that it not only allows to obtain a strong
lower bound for the concurrence without any numerical
optimization procedure, but also this bound is measurable, i.e., it is directly accessible in currently existing
laboratory experiments.

We will explain this in detail. We will see that, our bound is experimentally implementally by means of local observables, use the method as that of \cite{Mintert05PR,Mintert07,Huang09,Zhang08}.

First, note that the concurrence of a
bipartite pure state has another representation as \cite{Mintert05PR,Mintert07}
\begin{equation}\label{pure}
    C(|\psi\rangle)\equiv\sqrt{2(1-\mathrm{Tr}\rho_{A}^{2})}=
    \sqrt{\langle\psi|\otimes\langle\psi|A|\psi\rangle\otimes|\psi\rangle},
\end{equation}
where $A=4P_{-}^{(1)}\otimes P_{-}^{(2)}$.
$P_{-}^{(i)}$  is the projector on the antisymmetric
subspace $\mathcal{H}_{i}\wedge\mathcal{H}_{i}$ of the two copies of the
$i$th subsystem $\mathcal{H}_{i}\otimes\mathcal{H}_{i}$.Define $K_{1}=4P_{-}^{(1)}\otimes \eins^{(2)}$ and
$K_{2}=4(\eins^{(1)}\otimes P_{-}^{(2)})$, then $1-\tr\rho_{A}^{2}=\frac{1}{2}\tr(\rho\otimes\rho K_{1})$,
and $1-\tr\rho_{B}^{2}=\frac{1}{2}\tr(\rho\otimes\rho K_{2})$, so we can obtain
the term $\sqrt{(1-\mathrm{Tr}\rho_{A}^{2})(1-\mathrm{Tr}\rho_{B}^{2})}]$, provide that  we have two copies of the state, see \cite{Huang09,Zhang08}.

On the other hand, the term $\|\mathcal{R}(\rho-\rho_{A}\otimes\rho_{B})\|$ can be obtained by generalized entanglement witness \cite{HorodeckiPLA96,witness1}.Entanglement witnesses (EW) are Hermitian
operators that have positive averages on all separable states, but a negative one on at least one entangled state.It was shown that, a state is entangled if and only if it is detected by
some EW\cite{HorodeckiPLA96}. EW can be measured locally, and one can optimize such measurements in various aspects \cite{witness2}. Nowadays, entanglement witnesses are routinely used in experiments
to detect entanglement (see e.g., \cite{witness3}). Using the similar method of \cite{test}, we can directly measure the value of $\|\mathcal{R}(\rho-\rho_{A}\otimes\rho_{B})\|$ by a generalized witness. The method is as follows: from the result of matrix analysis\cite{horn}, every operator has a singular value decomposition (SVD), so we can get the SVD of $R(\rho-\rho_{A}\otimes \rho_{B})$ as $R(\rho-\rho_{A}\otimes \rho_{B})=UDV^{+}$, with $U,V$ unitary matrices, $D$ is a diagonal matrix. Then since the trace norm has a variational representation as $||A||=\max\{|\tr[X^{+}AY]|\}$, where $X,Y$ are unitary matrices, we can get that $\|\mathcal{R}(\rho-\rho_{A}\otimes\rho_{B})\|=\tr[VU^{+}\mathcal{R}(\rho-\rho_{A}\otimes\rho_{B})]$.
Now the only problem leaved is how to deal with the realignment operation. Note that every map $\phi$ on inner product space can induce its adjoint map $\phi^{\star}$,i.e., $\tr[\phi(X)Y^{+}]=\tr[X(\phi^{\star}(Y))^{+}]$, and for our question, it reads that $\tr[VU^{+}\mathcal{R}(\rho-\rho_{A}\otimes\rho_{B})]=\tr[\mathcal{R^{\star}}(VU^{+})(\rho-\rho_{A}\otimes\rho_{B})]$, and the adjoint map of the realignment operation is defined by $\mathcal{R^{\star}}(VU^{+}):=[\mathcal{R}^{-1}(VU^{+})^{T}]^{T}$, where $T$ is the transpose, and $\mathcal{R}^{-1}$ is the inverse map of the realignment operation. Now define $W:=[\mathcal{R}^{-1}(VU^{+})^{T}]^{T}$, then we get that
$\|\mathcal{R}(\rho-\rho_{A}\otimes\rho_{B})\|=\tr[W(\rho-\rho_{A}\otimes\rho_{B})]$, which is clearly physical accessible(e.g., see \cite{witness3}).

Next we consider some examples to illustrate further the tightness and
significance of our bound. To show that our bound is close to the real concurrence, note that in \cite{Ryu08}, the authors find a fast optimal algorithm to calculate the entanglement of formation of a mixed state, so we can use the optimal algorithm of \cite{Ryu08} to give the estimation of the concurrence, and comparing it with our bound. The concurrence using optimal method of \cite{Ryu08} is represented by blue colors, our low bound is represented by green colors.

{\bf Example 1} Isotropic states are a class of $U\otimes
U^{\ast }$ invariant mixed states in $d\times d$ systems
\begin{equation}
\rho _{F}={\frac{{1-F}}{d^{2}-1}}\left( I-|\Psi ^{+}\rangle \langle \Psi
^{+}|\right) +F|\Psi ^{+}\rangle \langle \Psi ^{+}|,  \label{isotropic}
\end{equation}%
where $|\Psi ^{+}\rangle \equiv \sqrt{1/d}\sum_{i=1}^{d}|ii\rangle $ and $%
F=\langle \Psi ^{+}|\rho _{F}|\Psi ^{+}\rangle $, satisfying $0\leq F\leq 1$%
, is the \emph{fidelity\/} of $\rho _{F}$ and $|\Psi ^{+}\rangle $. 

\begin{figure}[htbp]
\centering 
\includegraphics[width=0.4\textwidth]{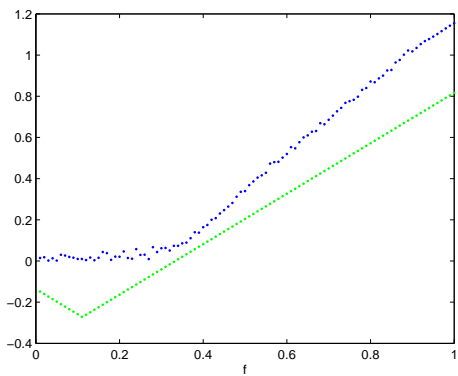}
\caption{Figure of example 1}
\end{figure}

{\bf Example 2.} Pawe{\l} Horodecki introduced a $3\times3$
bound entangled state in Ref. \cite{Horodeckibound}, and the density matrix
$\rho$ is real and symmetric,
\begin{equation}
\rho ={\frac{1}{8a+1}}\left(
\begin{array}{ccccccccc}
a & 0 & 0 & 0 & a & 0 & 0 & 0 & a \\
0 & a & 0 & 0 & 0 & 0 & 0 & 0 & 0 \\
0 & 0 & a & 0 & 0 & 0 & 0 & 0 & 0 \\
0 & 0 & 0 & a & 0 & 0 & 0 & 0 & 0 \\
a & 0 & 0 & 0 & a & 0 & 0 & 0 & a \\
0 & 0 & 0 & 0 & 0 & a & 0 & 0 & 0 \\
0 & 0 & 0 & 0 & 0 & 0 & {\frac{1+a}{2}} & 0 & {\frac{\sqrt{1-a^{2}}}{2}} \\
0 & 0 & 0 & 0 & 0 & 0 & 0 & a & 0 \\
a & 0 & 0 & 0 & a & 0 & {\frac{\sqrt{1-a^{2}}}{2}} & 0 &
{\frac{1+a}{2}}
\end{array}
\right), \label{rho}
\end{equation}
where $0<a<1$. Let us consider a mixture of this state with white
noise,
\begin{equation}
    \rho(p)=p\rho+(1-p)\frac{\eins}{9},
\end{equation}
and show the curves $1-\|\mathcal{R}(\rho)\|=0$,
$1-\|\tau\|-(\mathrm{Tr}\rho_{A}^{2}+ \mathrm{Tr}\rho_{B}^{2})/2=0$,
$\sqrt{(1-\mathrm{Tr}\rho_{A}^{2})(1-\mathrm{Tr}\rho_{B}^{2})}
    -\|\mathcal{R}(\rho-\rho_{A}\otimes\rho_{B})\|=0$ with respect to
the CCNR criterion, its optimal nonlinear witness, and Theorem 1 in
Fig. 2.It is found that the state $\rho(p)$ still has
entanglement when $p=0.9955$, $a=0.236$, using the CCNR criterion.
According to Theorem 1, one can obtain an upper bound $p=0.9939$,
$a=0.232$ for $\rho(p)$ which is still entangled.
\begin{figure}[htbp]
\centering 
\includegraphics[width=0.4\textwidth]{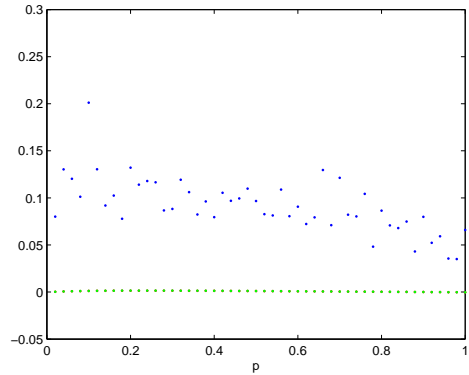}
\caption{Figure of example 2}
\end{figure}

When $p=0.9955$, $a=0.236$,using steepest descent method ,we get the concurrence is 0.101855,using our theorem ,the lower bound is 0.000487.
When $p=0.9939$, $a=0.232$,using steepest descent method ,we get the concurrence is 0.101758,using our theorem ,the lower bound is 0.000019.

{\bf Example 3.} Consider the following states introduced in Ref. \cite{Horodeckibound2}:
\begin{equation}
    \rho(\alpha)=\frac{2}{7}|\Psi ^{+}\rangle \langle \Psi
^{+}|+\frac{2}{7}\sigma_{+}+\frac{5-\alpha}{7}\sigma_{-},
\end{equation}
where $2\leq \alpha \leq 5$, $\sigma_{+}:=\frac{1}{3}(|0\rangle|1\rangle\langle 0|\langle 1|+|1\rangle|2\rangle\langle 1|\langle 2|+|2\rangle|0\rangle\langle 2|\langle 0|)$, $\sigma_{-}:=\frac{1}{3}(|1\rangle|0\rangle\langle 1|\langle 0|+|2\rangle|1\rangle\langle 2|\langle 1|+|0\rangle|2\rangle\langle 0|\langle 2|)$.

\begin{figure}[htbp]
\centering 
\includegraphics[width=0.4\textwidth]{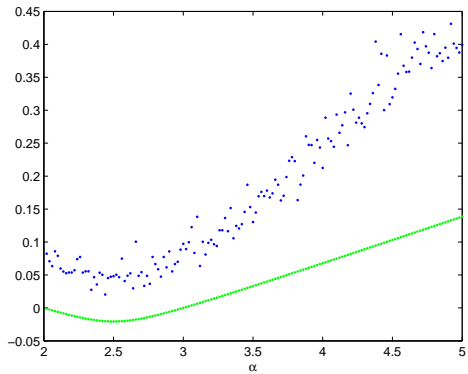}
\caption{Figure of example 3}
\end{figure}

In summary, we have provided an analytical
formula for a lower bound of concurrence, by finding a
connection with the currently most powerful detection criterion.
The bound is very close to the  actual values of concurrence for
some special class of quantum states. Also,
this bound is experimentally implementable and computationally very efficient, allowing to not only detect, but
also to quantify  entanglement in an experimental scenario.

\vskip 0.1 in {\noindent\bf Acknowledgment.} This work is supported
by NSF of China(10901103), partially supported by a grant of science
and technology commission of Shanghai Municipality (STCSM, No.
09XD1402500).

\end{document}